\documentclass[12pt]{spieman}  
\usepackage{amsmath,amsfonts,amssymb}
\usepackage{graphicx}
\usepackage{setspace}
\usepackage{tocloft}
\usepackage{multirow}
\usepackage{booktabs, multirow}
\usepackage[table]{xcolor}
\usepackage{caption} 
\usepackage{float}
\floatstyle{plaintop}
\usepackage{threeparttable}
\usepackage{subcaption}

\restylefloat{table}
\AtBeginDocument{}

\cftpagenumbersoff{figure}
\cftpagenumbersoff{table} 

\title{MR-Transformer: Vision Transformer for Total Knee Replacement Prediction Using Magnetic Resonance Imaging}

\author[a]{Chaojie Zhang}
\author[a]{Shengjia Chen}
\author[a]{Ozkan Cigdem}
\author[b]{Haresh Rengaraj Rajamohan}
\author[b]{Kyunghyun Cho}
\author[a]{Richard Kijowski}
\author[a]{Cem M. Deniz}

\affil[a]{Department of Radiology, New York University Langone Health, New York, NY, 10016, USA}
\affil[b]{Center for Data Science, New York University, New York, NY, 10011, USA}

\begin{document} 
\maketitle


\subsection*{\normalfont\textbf{Summary}}
A transformer-based deep learning model, MR-Transformer, was developed to predict total knee replacement using magnetic resonance imaging. The model exhibited state-of-the-art performance on the Osteoarthritis Initiative and Multicenter Osteoarthritis Study databases.


\subsection*{\normalfont\textbf{Key Points}}
\begin{itemize}
  \item MR-Transformer incorporates the ImageNet pre-training and captures three-dimensional spatial correlation to predict total knee replacement using MRI.
  \item The model achieved areas under the receiver operating characteristic curve (AUC) of 0.89 for coronal intermediate-weighted turbo spin-echo and 0.91 for sagittal intermediate-weighted turbo spin-echo with fat suppression from the Osteoarthritis Initiative database, as well as 0.82 for coronal short-tau inversion recovery and 0.82 for sagittal proton density fat-saturated from the Multicenter Osteoarthritis Study database.
  \item Using the coronal short-tau inversion recovery MRI knee scans, MR-Transformer outperformed other deep learning models in AUC (MR-Transformer: 0.82, TSE: 0.76 (\textit{P} $<$ .001), 3DMeT: 0.74 (\textit{P} = .005), and MRNet: 0.78 (\textit{P} = .010)). 
\end{itemize}

\subsection*{\normalfont\textbf{List of Abbreviations}}
\noindent 2D = two-dimensional, 3D = three-dimensional, OAI = Osteoarthritis Initiative, MOST = Multicenter Osteoarthritis Study, TKR = total knee replacement, AUC = area under receiver operating characteristic curve, COR IW TSE = coronal intermediate-weighted turbo spin-echo, SAG IW TSE FS = sagittal intermediated-weighted turbo spin-echo with fat suppression, COR STIR = coronal short-tau inversion recovery, SAG PD FAT SAT = sagittal proton density fat-saturated


\subsection*{\normalfont\textbf{Abstract}}

\noindent \textbf{Purpose:} To develop a transformer-based deep learning model that incorporates the ImageNet pre-training and captures three-dimensional (3D) spatial correlation to predict total knee replacement (TKR) using MRI.

\noindent \textbf{Materials and Methods:} A transformer-based deep learning model, MR-Transformer, was developed for TKR prediction. The model adapted from the ImageNet pre-trained vision transformer DeiT-Ti, directly utilizes 3D MR images as input. A total of 353 case-control matched pairs of coronal intermediate-weighted turbo spin-echo (COR IW TSE) and sagittal intermediated-weighted turbo spin-echo with fat suppression (SAG IW TSE FS) knee MR scans from the Osteoarthritis Initiative (OAI) and 270 case-control pairs of coronal short-tau inversion recovery (COR STIR) and sagittal proton density fat-saturated (SAG PD FAT SAT) knee MRI scans from the Multicenter Osteoarthritis Study (MOST) databases were utilized in the study. The performance of the proposed model was compared to existing state-of-the-art deep learning models for knee injury diagnosis using MRI, with the area under the receiver operating characteristic curve (AUC) as the evaluation metric.

\noindent \textbf{Results:} The proposed MR-Transformer model consistently achieved higher AUC performance compared to TSE, 3DMeT, and MRNet models for both coronal and sagittal knee MR scans from the OAI and MOST databases. Specifically, for COR STIR, MR-Transformer achieved a significantly higher AUC (0.82) compared to other DL models: TSE (0.76, \textit{P} $<$ .001), 3DMeT (0.74, \textit{P} = .005), and MRNet (0.78, \textit{P} = .010). Similarly, for COR IW TSE, SAG IW TSE FS, and SAG PD FAT SAT, MR-Transformer achieved AUCs of 0.89, 0.91, and 0.82, respectively. These were significantly higher than those of TSE (0.86, \textit{P} = .009; 0.82, \textit{P} $<$ .001; 0.75, \textit{P} = .006) and 3DMeT (0.82, \textit{P} = .004; 0.78, \textit{P} $<$ .001; 0.63, \textit{P} $<$ .001), respectively. While the proposed model achieved either similar or slightly higher AUCs than MRNet (0.89, 0.90, and 0.81 for COR IW TSE, SAG IW TSE FS, and SAG PD FAT SAT, respectively), the differences were not statistically significant (\textit{P} = .44, .070, and .20, respectively).

\noindent \textbf{Conclusion:} MR-Transformer exhibited state-of-the-art performance on TKR prediction using MRI compared to currently available deep learning models.

\begin{spacing}{1.2}   
\section{Introduction}
\label{sect:intro}  

Knee osteoarthritis, the most prevalent type of arthritis, is a degenerative joint disease that significantly diminishes the quality of life for millions worldwide, resulting in pain, limited mobility, and disability \cite{kellgren1957radiological, altman1986development}. While knee osteoarthritis has no cure, total knee replacement (TKR) offers a viable treatment option for advanced stages of the disease \cite{felson1998update}. Early identification of patients at risk of TKR progression is crucial to facilitate the development and implementation of potential disease-modifying therapies \cite{felson1998update}.
Knee osteoarthritis is also a structural disease that results in structural changes within the joint, the diagnosis of it requires comprehensive inspection of imaging studies like MRI, by an experienced radiologist. Recently, deep learning methods have shown the potential to assist in TKR prediction \cite{rajamohan2023prediction,tolpadi2020deep,leung2020prediction}.


Several challenges arise when employing deep learning methods for predicting TKR from MRI data: the three-dimensional (3D) nature of the data, the limited size of medical datasets, and the inclusion of global structural information pertinent to knee osteoarthritis.
Current studies implemented 3D convolutional neural networks to capture 3D spatial information on the MR images to predict TKR\cite{rajamohan2023prediction,tolpadi2020deep}.
However, deep learning as a data-driven method, often requires a large amount of data to achieve optimal performance. Deep learning models without any pre-training usually achieve suboptimal performance on small medical datasets\cite{matsoukas2021time}. To address this problem, MRNet used the ImageNet pre-trained two-dimensional (2D) convolutional neural network to encode each MR slice and aggregated all encoded features to diagnose knee injuries from knee MRI scans\cite{bien2018deep}. This approach leverages large-scale pre-training but fails to capture the 3D spatial correlation of MR images.

In this study, we proposed MR-Transformer, a novel deep learning model adapted from the ImageNet pre-trained vision transformer DeiT-Ti\cite{touvron2021training}, for TKR prediction using MRI. The model inherits the ImageNet\cite{deng2009imagenet} pre-trained weights and captures 3D spatial correlation of MR images. In addition, the transformer architecture modeling long-range dependencies enables the model to better capture global structural information from the MR images. We evaluated our proposed model on TKR prediction using MR images from the Osteoarthritis Initiative (OAI)\cite{peterfy2008osteoarthritis} and Multicenter Osteoarthritis Study (MOST)\cite{segal2013multicenter} databases. MR-Transformer exhibited state-of-the-art performance on TKR prediction using MRI.

\section{Materials and Methods}
\subsection{Data}
In this study, we used MR images from two publicly available databases established as prospective cohort studies: OAI\cite{peterfy2008osteoarthritis} and MOST\cite{segal2013multicenter}. The OAI database contains MRI examinations of different tissue contrasts from 4796 subjects with or at risk for knee osteoarthritis evaluated at baseline and 12, 24, 36, 48, 60, 72, 84, and 108-month follow-up. The MOST database contains MRI examinations from 3026 subjects with or at risk for knee osteoarthritis evaluated at baseline and 15, 30, 60, and 84-month follow-up. The OAI and MOST databases were approved by the Internal Review Boards at University of California at San Francisco, Boston University Medical Center, and each individual clinical recruitment site and was performed in compliance with the Declaration of Helsinki. All subjects signed written informed consent. In both databases, balanced case–control cohorts were selected by matching case subjects and control subjects using baseline demographic variables associated with knee osteoarthritis progression including age, sex, ethnicity, and body mass index. Case subjects were identified as individuals who underwent a TKR after the baseline enrollment date, while control subjects were identified as individuals who did not. 

In the OAI database, a total of 353 case–control pairs were identified and separated into the training group of 302 pairs and the test group of 51 pairs. The deep learning models were trained on coronal intermediate-weighted turbo spin-echo (COR IW TSE) and sagittal intermediated-weighted turbo spin-echo with fat suppression (SAG IW TSE FS) MRI knee scans of the training group using six-fold cross-validation.

In the MOST database, a total of 270 case–control pairs were identified and separated into the training group of 231 pairs and the test group of 39 pairs. The deep learning models were trained on coronal short-tau inversion recovery (COR STIR) and sagittal proton density fat-saturated (SAG PD FAT SAT) MRI knee scans of the training group using six-fold cross-validation.

\subsection{Model Development}
The proposed model, MR-Transformer, was adapted from the ImageNet pre-trained tiny vision transformer DeiT-Ti\cite{touvron2021training}. The main idea of the adaptation is to modify the 2D vision transformer to accommodate 3D input while retaining the pre-trained model weights.

Different from traditional convolutional neural networks, vision transformers separate the images into small 2D patches and encode the vectorized patches using self-attention mechanisms. This patch-based processing enables vision transformers to be flexible in handling images with different dimensions, making the adaptation from 2D inputs to 3D inputs possible. In our adaptation, we split the 3D MR images into 2D patches, retain the pre-trained model weights, and modify the position embeddings of the model.

\begin{figure}[!ht]
\centering
\includegraphics[width=0.8\textwidth]{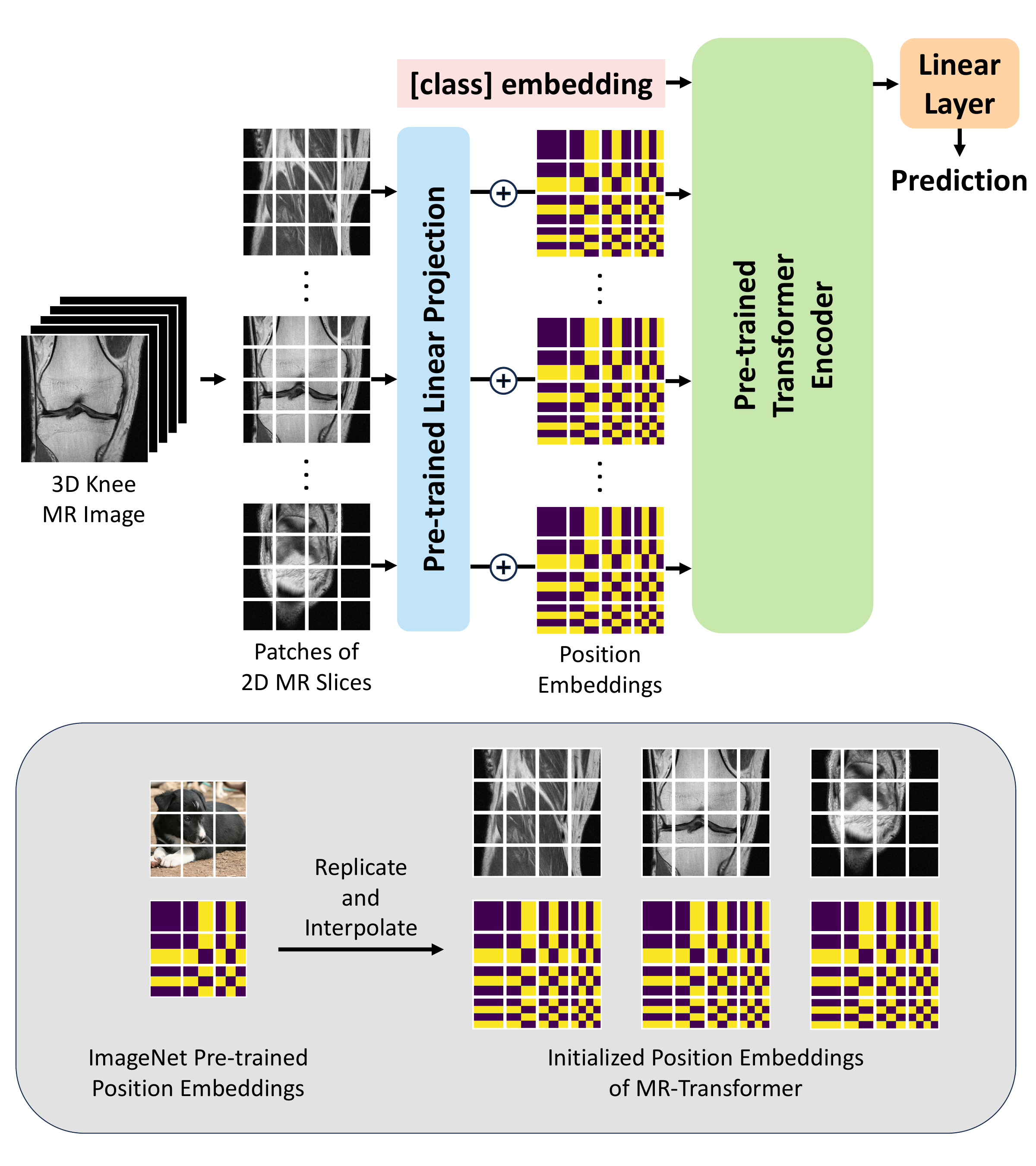}
\caption{Model architecture of MR-Transformer. The model inherited the model weights of the ImageNet pre-trained vision transformer and processed knee MR images by separating them into small 2D patches.}
\label{Flowchart}
\end{figure}

The model architecture of MR-Transformer is provided in Figure \ref{Flowchart}. In the pipeline of MR-Transformer, the single grayscale channel of the MR image is replicated to generate a three-color-channel input matrix. The 3D input matrix is then separated into 16$\times$16 patches, with each patch being flattened into a 768-dimensional vector. Then, the linear projection layer inherited from DeiT-Ti project the flattened patch vectors into 192-dimensional patch embeddings. The adapted learnable position embeddings are added to the corresponding patch embeddings to retain the positional information of each 2D patch. The patch embeddings, appended with an inherited learnable "class" embedding, are processed by the pre-trained transformer encoder from DeiT-Ti. Then the final linear layer utilizes the encoded "class" embedding to generate the TKR prediction outcome.
 
 MR-Transformer was designed to capture 3D spatial correlation of MR images, meanwhile leveraging the advantages of 2D large-scale pre-training. Moreover, the transformer architecture of MR-Transformer enabled the modeling of long-range dependencies, facilitating the capture of global structural information from the MR images.


\subsection{Position Embeddings}
The key components of MR-Transformer can be easily inherited from the ImageNet pre-trained DeiT-Ti except for the position embeddings. Position embeddings are learnable vectors with the same dimension as the patch embeddings. They are used to retain the positional information of different patches. To adapt the 2D pre-trained position embeddings to the 3D MR images, we replicated the 2D pre-trained position embeddings for each MR slice. Then, 2D interpolation was employed to the position embeddings when the MR slice had a larger size than the pre-training images. Note that position embeddings are learnable vectors that can be trained to learn the 3D spatial position information of MR images.


\subsection{Comparison with Other Models}
The proposed MR-Transfomer model was compared to three different deep learning models designed for knee injury diagnosis from MRI. MRNet is a convolutional neural network-based model developed for detecting general abnormalities and specific diagnoses on knee MRI exams, it employs the ImageNet pre-trained AlexNet\cite{krizhevsky2012imagenet} to encode each MR slice and uses max pooling to aggregate the encoded features\cite{bien2018deep}. TSE is a 3D convolutional neural network model developed for TKR prediction, it captures 3D spatial correlation from knee MR images\cite{rajamohan2023prediction}. 3DMeT is a 3D transformer-based model with a convolutional neural network teacher mechanism, it was developed for knee cartilage defect assessment from knee MRI\cite{wang20213dmet}.

\subsection{Statistical Analysis}
Receiver operator characteristic analysis for area under the curve (AUC) and the calculation of 95\% confidence interval were employed for individual model evaluation. Sensitivity and specificity assessments were conducted at high specificity (80\%) and sensitivity (80\%) thresholds, respectively. 
Statistical analyses were implemented utilizing the R Project version 4.2.1 for Statistical Computing Software\cite{r2013r}. The one-sided paired t-tests were used to evaluate differences in performance between the different models. Statistical significance was defined as a \textit{P} value less than 0.05.

\section{Results}
\subsection{Participant Characteristics}
\begin{figure}[hbt!]
\centering
\begin{subfigure}{0.45\textwidth}
    \includegraphics[width=\linewidth]{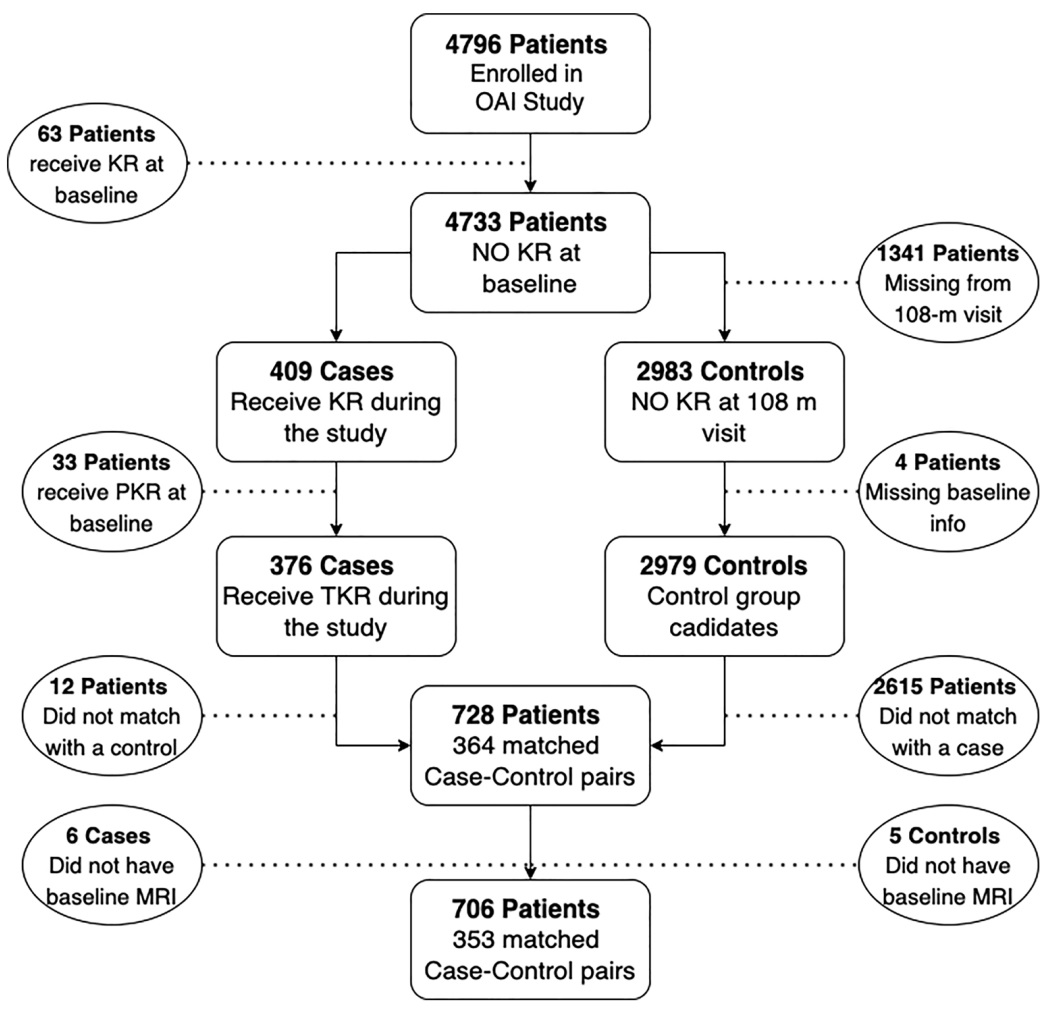}
    \caption{}
    \label{DatasetDistributionsOAI}
\end{subfigure}
    \hspace{1cm} 
\begin{subfigure}{0.45\textwidth}
    \includegraphics[width=\linewidth]{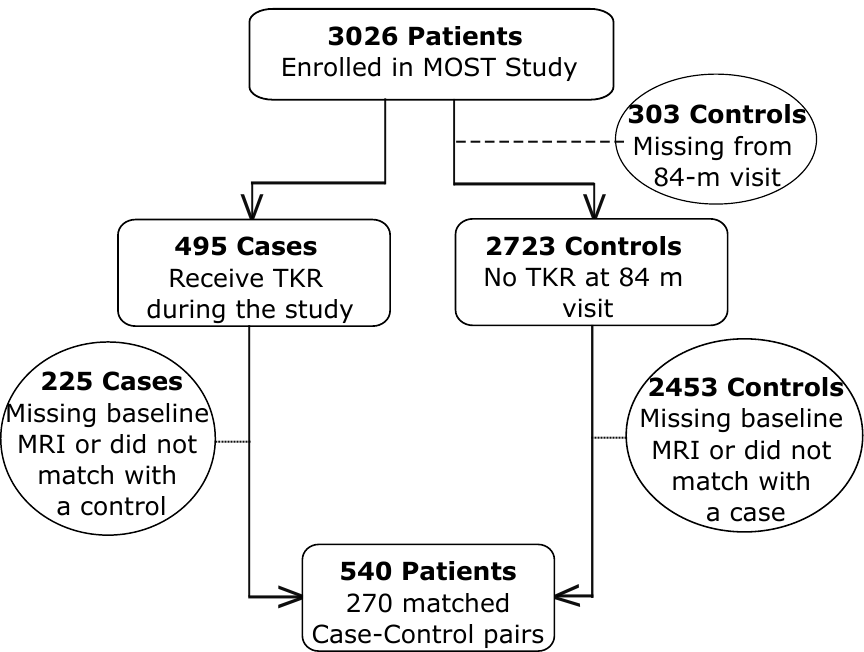}
    \caption{}
    \label{DatasetDistributionsMOST}
\end{subfigure}
\caption{Selection of the study cohorts. a) 353 case–control pairs of subjects in the OAI database, b) 270 case–control pairs of subjects in the MOST database.}
\label{DatasetDistributions}
\end{figure}

A total of 353 case–control pairs were derived from the 4796 subjects in the OAI database, while 270 case–control pairs were derived from the 3026 subjects in the MOST database. Subjects were excluded if they had TKR at baseline, received partial knee replacement over the course of follow-up, were missing baseline or 108-month follow-up information, or did not match with a case or control subject. The flowcharts of selecting case-control pairs from the OAI and MOST databases are provided in Figure \ref{DatasetDistributions}. In the 353 case–control pairs, there were 138 males aged from 45 to 78 with a mean$\pm$standard deviation of 64$\pm$8 and 215 females aged from 45 to 79 with a mean$\pm$standard deviation of 63$\pm$8. In 270 case–control pairs, there were 67 males aged from 50 to 78 with a mean$\pm$standard deviation of 65$\pm$7 and 203 females aged from 50 to 79 with a mean$\pm$standard deviation of 65$\pm$7. The baseline demographics of study cohorts are provided in Table \ref{tab: demographicInfo}.

\begin{table}[!htp]\centering
\caption{Summary Statistics of Demographic Variables for Matched Case-Control Cohorts}\label{tab: demographicInfo}
\scriptsize

\begin{tabular}{ccccclccl}
\toprule
\multirow{2}{*}{Dataset} & \multicolumn{2}{c}{\multirow{2}{*}{Parameter}} & \multicolumn{2}{c}{Men}          & \multicolumn{1}{c}{\multirow{2}{*}{P Value}} & \multicolumn{2}{c}{Women}        & \multicolumn{1}{c}{\multirow{2}{*}{P Value}} \\ \cmidrule{4-5} \cmidrule{7-8}
                         & \multicolumn{2}{c}{}                           & Patients      & Control Patients & \multicolumn{1}{c}{}                         & Patients      & Control Patients & \multicolumn{1}{c}{}                         \\ \midrule
\multirow{10}{*}{OAI}    & \multicolumn{2}{c}{No. of patients}            & 138           & 138              &                                              & 215           & 215              &                                              \\  
                         & \multicolumn{2}{c}{Age range (y)}              & 45-78         & 45-78            &                                              & 45-79         & 45-79            &                                              \\  
                         & \multicolumn{2}{c}{Mean age (y)}               & 64$\pm$8      & 64$\pm$8         & \multicolumn{1}{c}{\textgreater{}0.99}       & 63$\pm$8      & 63$\pm$8         & \multicolumn{1}{c}{\textgreater{}0.99}       \\ 
                         & \multicolumn{2}{c}{Mean height (m)}            & 1.76$\pm$0.06 & 1.76$\pm$0.06    & \multicolumn{1}{c}{0.78}                     & 1.62$\pm$0.06 & 1.62$\pm$0.06    & \multicolumn{1}{c}{0.56}                     \\ 
                         & \multicolumn{2}{c}{Mean weight (kg)}           & 93.4$\pm$14.1 & 92.0$\pm$12.7    & \multicolumn{1}{c}{0.38}                     & 78.5$\pm$14.8 & 77.0$\pm$13.6    & \multicolumn{1}{c}{0.28}                     \\  
                         & \multicolumn{2}{c}{Mean BMI (kg/m2)}           & 29.9$\pm$3.8  & 29.4$\pm$3.4     & \multicolumn{1}{c}{0.31}                     & 29.9$\pm$5.3  & 29.5$\pm$4.7     & \multicolumn{1}{c}{0.41}                     \\ \cmidrule{2-9} 
                         & \multirow{4}{*}{Ethnicity*} & White            & 126           & 126              &                                              & 177           & 177              &                                              \\ 
                         &                             & African American & 10            & 10               &                                              & 33            & 33               &                                              \\  
                         &                             & Asian            & 0             & 0                &                                              & 2             & 2                &                                              \\  
                         &                             & Other nonwhite   & 2             & 2                &                                              & 3             & 3                &                                              \\ \midrule
\multirow{10}{*}{MOST}   & \multicolumn{2}{c}{No. of patients}            & 67            & 67               &                                              & 203           & 203              &                                              \\  
                         & \multicolumn{2}{c}{Age range (y)}              & 50-78         & 50-78            &                                              & 50-79         & 50-79            &                                              \\  
                         & \multicolumn{2}{c}{Mean age (y)}               & 65$\pm$7      & 65$\pm$7         & \multicolumn{1}{c}{0.99}                     & 65$\pm$7      & 65$\pm$7         & \multicolumn{1}{c}{0.99}                     \\  
                         & \multicolumn{2}{c}{Mean height (m)}            & 1.79$\pm$0.06 & 1.78$\pm$0.06    & \multicolumn{1}{c}{0.46}                     & 1.62$\pm$0.06 & 1.63$\pm$0.06    & \multicolumn{1}{c}{0.41}                     \\  
                         & \multicolumn{2}{c}{Mean weight (kg)}           & 94.8$\pm$13.0 & 96.7$\pm$18.2    & \multicolumn{1}{c}{0.61}                     & 81.6$\pm$13.6 & 81.6$\pm$12.6    & \multicolumn{1}{c}{0.99}                     \\  
                         & \multicolumn{2}{c}{Mean BMI (kg/m2)}           & 30.1$\pm$4.2  & 30.0$\pm$4.1     & \multicolumn{1}{c}{0.98}                     & 31.1$\pm$4.9  & 30.9$\pm$4.7     & \multicolumn{1}{c}{0.71}                     \\ \cmidrule{2-9} 
                         & \multirow{4}{*}{Ethnicity*} & White            & 67            & 67               &                                              & 188           & 188              &                                              \\  
                         &                             & African American & 0             & 0                &                                              & 15            & 15               &                                              \\  
                         &                             & Asian            & 0             & 0                &                                              & 0             & 0                &                                              \\  
                         &                             & Other nonwhite   & 0             & 0                &                                              & 0             & 0                &                                              \\ \bottomrule
\end{tabular}

\begin{tablenotes}
\item Note: Mean data are presented as mean$\pm$standard deviation. The \textit{P}-value compares the differences in mean between case and control groups for each variable. BMI = Body Mass Index, MOST= Multi-Center Osteoarthritis Study, OAI= Osteoarthritis Initiative. *Number of patients are shown.
\end{tablenotes}
\end{table}

\subsection{Comparison of Models}
The comparison of models for TKR prediction is presented in Table \ref{tab: model performance}. When using MR images of different contrasts from the OAI and MOST databases, MR-Transformer attained significantly higher AUCs (COR IW TSE: 0.89 and SAG IW TSE: 0.91 and COR STIR: 0.82 and SAG PD FAT SAT: 0.82) than TSE (COR IW TSE: 0.86, \textit{P} = .009 and SAG IW TSE: 0.82, \textit{P} $<$ .001 and COR STIR: 0.76, \textit{P} $<$ .001 and SAG PD FAT SAT: 0.75, \textit{P} = .006) and 3DMeT (COR IW TSE: 0.82, \textit{P} = .004 and SAG IW TSE: 0.78, \textit{P} $<$ .001 and COR STIR: 0.74, \textit{P} = .005 and SAG PD FAT SAT: 0.63, \textit{P} $<$ .001). Similarly, using COR STIR MR images from the MOST database, MR-Transformer attained significantly higher AUC than MRNet (0.78, \textit{P} = .010). However, the utilization of MR images with other tissue contrasts did not yield significant differences in AUCs between MR-Transformer and MRNet (COR IW TSE: 0.89, \textit{P} = .44 and SAG IW TSE: 0.90, \textit{P} = .070 and SAG PD FAT SAT: 0.81, \textit{P} = .20).

Using COR IW TSE (OAI) MR scans, the MR-Transformer achieved a significantly higher sensitivity (0.84) compared to TSE (0.75, \textit{P} = .013) and 3DMeT (0.68, \textit{P} = .014). While the sensitivity of the MR-Transformer was also higher than MRNet (0.80),  this difference was not statistically significant (\textit{P} = .10). Similarly, the MR-Transformer's specificity (0.80) was significantly higher than TSE (0.74, \textit{P} = .012) and 3DMeT (0.65, \textit{P} = .010), yet not statistically different from MRNet (0.78, \textit{P} = .25). For SAG IW TSE (OAI), the MR-Transformer achieved a sensitivity of 0.83, which was significantly higher than TSE (0.68, \textit{P} $<$ .001) and 3DMeT (0.64, \textit{P} $<$ .001). Although the sensitivity of MR-Transformer was lower than MRNet (0.84), the difference was not statistically significant (\textit{P} = .74). Similarly, the MR-Transformer's specificity (0.84) was significantly higher than TSE (0.69, \textit{P} = .001) and 3DMeT (0.55, \textit{P} $<$ .001), yet not statistically different from MRNet (0.83, \textit{P} = .33).

Using COR STIR (MOST) MR images, the MR-Transformer achieved a significantly higher sensitivity (0.62) compared to 3DMeT (0.53, \textit{P} = .037). Although the sensitivity of the MR-Transformer was also higher than TSE (0.59) and MRNet (0.57), these differences were not statistically significant (\textit{P} = .17 and \textit{P} = .075, respectively). Conversely, the specificity of MR-transformer (0.66) was significantly higher than TSE (0.47, \textit{P} = .001), 3DMeT (0.49, \textit{P} $<$ .001), and MRNet (0.59, \textit{P} = .032). For  SAG PD FAT SAT (MOST), the MR-Transformer achieved a sensitivity of 0.64, which was significantly higher than TSE (0.52, \textit{P} = .030) and 3DMeT (0.38, \textit{P} = .002). Although the sensitivity of MR-Transformer was also higher than MRNet (0.62), the difference was not statistically significant (\textit{P} = .16). Similarly, the MR-Transformer's specificity (0.58) was significantly higher than 3DMeT (0.30, \textit{P} $<$ .001). Even though the sensitivity of MR-Transformer was higher than 3DMeT (0.49) and lower than MRNet (0.60), these differences were not statistically significant (\textit{P} = .053 and \textit{P} = .68, respectively).

\begin{table}[!htp]\centering
\caption{Comparison of Deep Learning Models for Total Knee Replacement Prediction.}\label{tab: model performance}
\scriptsize

\begin{tabular}{cccccccc}
\toprule
                                & Model          & AUC                    & P Value         & \begin{tabular}[c]{@{}c@{}}Sensitivity \\ (at 80\% Specificity)\end{tabular} & P Value         & \begin{tabular}[c]{@{}c@{}}Specificity \\ (at 80\% Sensitivity)\end{tabular} & P Value         \\ \midrule
\multirow{4}{*}{COR IW TSE}     & TSE            & 0.86 ± 0.02          & .009            & 0.75 ± 0.07                                                                & .013            & 0.74 ± 0.04                                                                & .012            \\
                                & 3DMeT          & 0.82 ± 0.03          & .004            & 0.68 ± 0.07                                                                & .014            & 0.65 ± 0.08                                                                & .010            \\
                                & MRNet          & 0.89 ± 0.01          & .44             & 0.80 ± 0.01                                                                & .10             & 0.78 ± 0.03                                                                & .25             \\
                                & MR-Transformer & \textbf{0.89 ± 0.01} & $*$             & \textbf{0.84 ± 0.04}                                                       & $*$            & \textbf{0.80 ± 0.03}                                                       & $*$             \\ \midrule
\multirow{4}{*}{SAG IW TSE}     & TSE            & 0.82 ± 0.01          & \textless{}.001 & 0.68 ± 0.03                                                                & \textless{}.001 & 0.69 ± 0.02                                                                & .001            \\
                                & 3DMeT          & 0.78 ± 0.02          & \textless{}.001 & 0.64 ± 0.03                                                                & \textless{}.001 & 0.55 ± 0.08                                                                & \textless{}.001 \\
                                & MRNet          & 0.90 ± 0.00          & .070            & \textbf{0.84 ± 0.02}                                                       & .74             & 0.83 ± 0.02                                                                & .33             \\
                                & MR-Transformer & \textbf{0.91 ± 0.01} & $*$             & 0.83 ± 0.02                                                                & $*$            & \textbf{0.84 ± 0.03}                                                       & $*$             \\ \midrule
\multirow{4}{*}{COR STIR}       & TSE            & 0.76 ± 0.01          & \textless{}.001 & 0.59 ± 0.04                                                                & .17             & 0.47 ± 0.06                                                                & .001            \\
                                & 3DMeT          & 0.74 ± 0.03          & .005            & 0.53 ± 0.09                                                                 & .037            & 0.49 ± 0.06                                                                & \textless{}.001 \\
                                & MRNet          & 0.78 ± 0.01          & .010            & 0.57 ± 0.07                                                                & .075            & 0.59 ± 0.03                                                                 & .032            \\
                                & MR-Transformer & \textbf{0.82 ± 0.01} & $*$             & \textbf{0.62 ± 0.04}                                                       & $*$            & \textbf{0.66 ± 0.03}                                                       & $*$             \\ \midrule
\multirow{4}{*}{SAG PD FAT SAT} & TSE            & 0.75 ± 0.02          & .006            & 0.52 ± 0.08                                                                & .030            & 0.49 ± 0.05                                                                & .053            \\
                                & 3DMeT          & 0.63 ± 0.03          & \textless{}.001 & 0.38 ± 0.07                                                                 & .002            & 0.30 ± 0.05                                                                 & \textless{}.001 \\
                                & MRNet          & 0.81 ± 0.01          & .20             & 0.62 ± 0.03                                                                & .16             & \textbf{0.60 ± 0.02}                                                       & .68             \\
                                & MR-Transformer & \textbf{0.82 ± 0.02} & $*$             & \textbf{0.64 ± 0.04}                                                       & $*$            & 0.58 ± 0.07                                                                & $*$             \\ \bottomrule
\end{tabular}
\begin{tablenotes}
\item Note: The MR-Transformer model was compared against three baseline deep learning models based on AUC, sensitivity at 80\% specificity, and specificity at 80\% sensitivity. The metrics (AUC, sensitivity, specificity) are presented as mean ± 95\% confidence interval, calculated from six cross-validation folds. AUC = area under receiver operating characteristic curve, COR IW TSE = coronal intermediate-weighted turbo spin-echo, SAG IW TSE FS = sagittal intermediated-weighted turbo spin-echo with fat suppression, COR STIR = coronal short-tau inversion recovery, SAG PD FAT SAT = sagittal proton density fat-saturated, $*$: the reference model for one-sided paired t-test. 
\end{tablenotes}
\end{table}

\subsection{Model Interpretation}
\begin{figure}[!ht]
\centering
\includegraphics[width=0.8\textwidth]{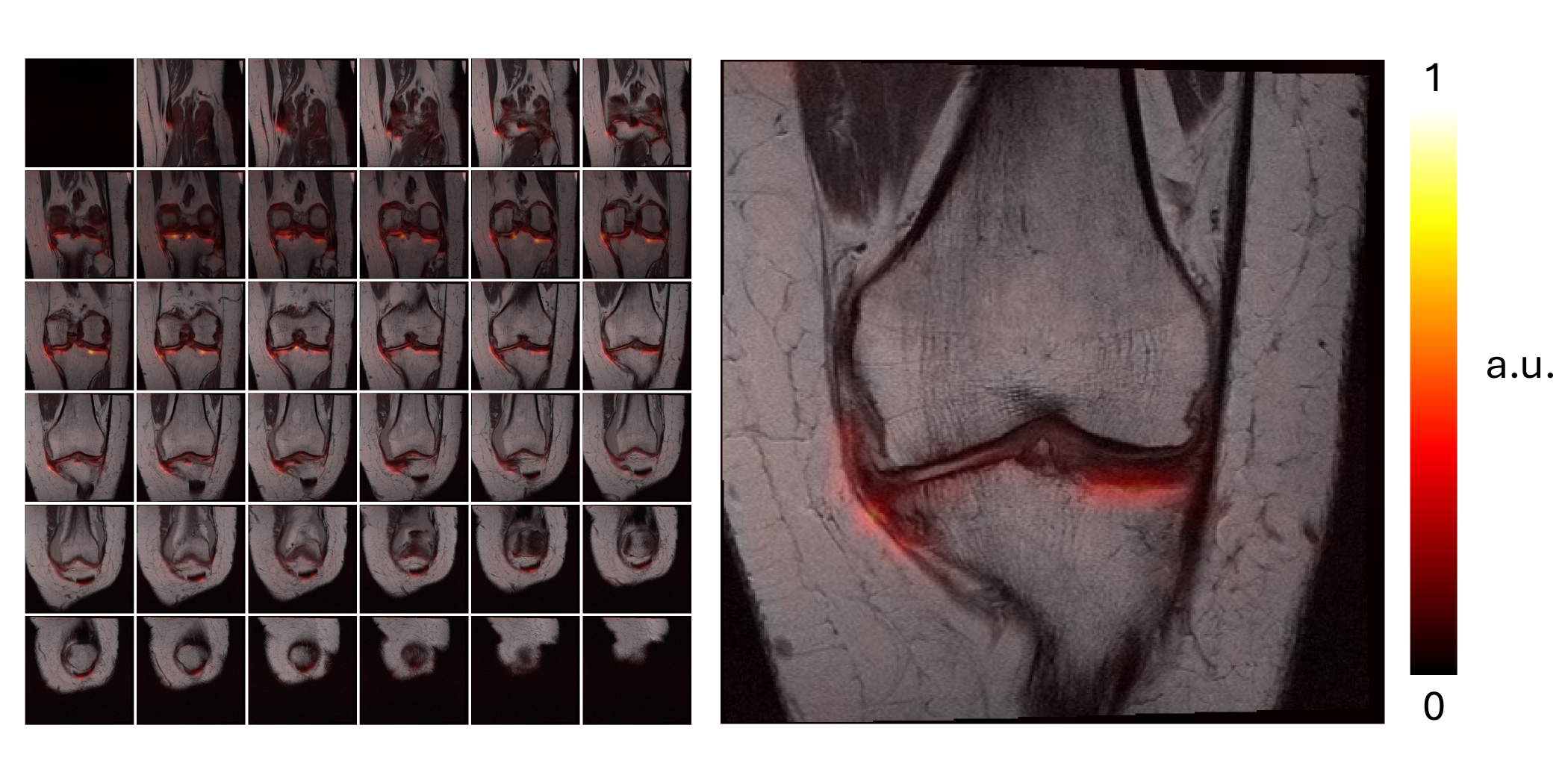}
\caption{Example attention map for a knee MR image in the test set from a subject who underwent a TKR within nine years. Highlighted regions indicate areas where the MR-Transformer model focuses to make decisions regarding the probability of TKR within nine years.}
\label{AttentionMap}
\end{figure}

Attention maps generated using Attention Rollout \cite{abnar2020quantifying} can highlight crucial regions within MR images. Figure~\ref{AttentionMap} presents an MR image from a subject who underwent a TKR within nine years. The highlighted area in the joint region indicates informative regions from the MR image relevant to TKR prediction.

\section{Discussion}
In this study, we introduced MR-Transformer, a transformer-based deep learning model designed for predicting TKR using sagittal and coronal MRI scans from the OAI and MOST databases. The proposed model exhibits several advantages, including leveraging large-scale ImageNet pre-training, capturing 3D spatial correlation, and modeling long-range dependencies from MR images. For TKR prediction using COR STIR knee MRI scans from the MOST database, MR-Transformer achieved a significantly higher AUC (0.82) compared to other deep learning models: TSE (0.76, \textit{P} $<$ .001), 3DMeT (0.74, \textit{P} = .005), and MRNet (0.78, \textit{P} = .010). Similarly, MR-Transformer also achieved significantly higher AUCs compared to TSE and 3DMeT for other tissue contrasts in MRI scans: COR IW TSE (0.89 vs 0.86, \textit{P} = .009 and 0.82, \textit{P} = .004), SAG IW TSE (0.91 vs 0.82, \textit{P} $<$ .001 and 0.78, \textit{P} $<$ .001), and SAG PD FAT SAT (0.82 vs 0.75, \textit{P} = .006 and 0.63, \textit{P} $<$ .001), while differences compared to MRNet were not statistically significant for these contrasts.


Compared with TSE and 3DMeT, MR-Transformer leveraged ImageNet pre-trained weights and achieved better performance in TKR prediction, while TSE and 3DMeT are models trained from scratch. The results demonstrate the importance of large-scale pre-training in small medical datasets. As MRNet also leveraged ImageNet pre-training, it achieved close performance to MR-Transformer in TKR prediction. However, we still observed a better performance of MR-Transformer when using COR STIR MR images. This could be attributed to MR-Transformer's capability to capture 3D spatial correlation and model long-range dependencies from MR images.

Although MR-Transformer exhibited leading performance in TKR prediction using MRI, some limitations of the model must be noted. The computation burden of MR-Transformer is heavy especially when processing large MR images. Since the self-attention mechanism of transformer models has quadratic computation complexity to the number of input patches, the considerable number of separated patches from a 3D MR image would bring an overwhelming computation load (a $36\times512\times512$ MRI matrix would be separated into $36,864$ patches). It is recommended to use small MR images when the computational resources are limited. As a future work, exploring the extension of MR-Transformer to diverse MRI classification tasks can be considered, aiming to investigate its generalizability and potential adaptations for various disease diagnoses.

The application of deep learning models in disease diagnosis signals the potential of artificial intelligence in medical fields. The proposed MR-Transformer, incorporating ImageNet pre-training and capturing three-dimensional spatial correlation, predicts TKR using MRI. It achieved superior AUC performance compared to TSE, 3DMeT, and MRNet across coronal and sagittal knee MR scans from the OAI and MOST databases, demonstrating its potential for TKR prediction. The open-source implementation of our method is available at \href{https://github.com/denizlab/MR-Transformer}{https://github.com/denizlab/MR-Transformer}.

\newpage
\bibliography{reference}   
\bibliographystyle{spiejour}   

\end{spacing}

\end{document}